\documentclass{mem}
\usepackage{natbib}\usepackage{txfonts}\usepackage{balance}
\usepackage{flushend}
\usepackage{graphicx}
\usepackage[breaklinks,pdftex]{hyperref}
\idline{75}{282}
\begin{document}
\def\teff{$T\rm_{eff }$}
\def\kms{$\mathrm {km s}^{-1}$}

\title{
Particle acceleration by turbulent-driven magnetic reconnection and the production of gamma-rays and neutrinos in AGNs 
}

   \subtitle{}

\author{
E. M. de Gouveia Dal Pino\inst{1}
\and J. C. Ramirez-Rodriguez\inst{2}
\and T. Medina-Torrejon\inst{3}
\and G. H. Vicentin\inst{1}
\and L. Passos Reis\inst{1}
}

\institute{
IAG-USP, Instituto de Astronomia, Geofísica e Ciências Atmosféricas  --
Universidade de São Paulo, R. do Matão 1226, São Paulo, SP,  05508-090, Brazil
\and
CBPF, Centro Brasileiro de Pesquisas Físicas, Rio de Janeiro, Brazil
\and
IFSC-USP, Instituto de Física de São Carlos, Universidade de São Paulo, 
Brazil
}

\authorrunning{de Gouveia Dal Pino et al.}

\titlerunning{Turbulent-driven Magnetic Reconnection and Gamma-Rays }

\date{Received: XX-XX-XXXX (Day-Month-Year); Accepted: XX-XX-XXXX (Day-Month-Year)}

\abstract{
3D Magnetohydrodynamic (MHD) resistive simulations have highlighted the significance of ubiquitous turbulence to drive fast reconnection. It has been demonstrated that particle acceleration via reconnection in 3D magnetized flows, where turbulence is embedded in large-scale magnetic fields such as in relativistic jets and accretion flows around compact sources, is remarkably efficient. Particles experience Fermi acceleration across all scales of turbulent reconnection layers, outweighing the considerably slower drift acceleration mechanism. This stands in contrast to recent assertions stemming from  PIC simulations, that claimed the dominance of the latter process. In this talk, I review how particle acceleration is driven by 3D turbulent reconnection to very high energies and demonstrate its potential in magnetized regions of AGN jets and accretion disks to explain the gamma-ray and neutrino emissions. Applications to sources such as TXS 0506+056, MRK 501, and NGC 1068 are discussed.

}
\maketitle{}

\section{Introduction}

Recent advances in high-energy astrophysics emphasize the critical role of magnetic reconnection in accelerating energetic particles and generating  very high-energy (VHE) flares in magnetically dominated regions of compact systems like accretion flows, relativistic jets, pulsar wind nebulae, and gamma-ray bursts \citep[e.g.,][]{dalpino_lazarian_2005,giannios_etal_09,zhang_yan_11,cerutti_etal_2013,kadowaki_etal_15,petropoulou_etal_2016,lyutikov_etal_2018,
Medina-Torrejón_2021}.  

Studies of particle acceleration driven by magnetic reconnection span two primary scales: kinetic (micro) scales explored mainly through 2D particle-in-cell (PIC) simulations \citep[e.g.,][]{
drake_etal_2006,cerutti_etal_2013,werner_etal_2018,
lyutikov_etal_2018,
sironi_spitkovsky_2014,guo_etal_2016,comisso_sironi_2021,zhang2021}, and macroscopic astrophysical scales addressed through 3D MHD
simulations \citep[e.g.,][]{kowal_etal_2011,kowal_etal_2012,dalpino_kowal_15,lazarian12,delvalle_etal_16,beresnyak_etal_2016,
Medina-Torrejón_2021,Medina-Torrejón_2023, Zhang_Xu2023}.  

A review of the similarities and differences between these two scales can be found in \citealt{dalpino_torrejon2024}. They  can be succinctly summarized as follows.


PIC simulations probe scales ranging from 100 to few 1000 times the plasma inertial length ($c/\omega_{\rm p}$, where $c$ is the speed of light and $\omega_{\rm p}$  the plasma frequency). These scales correspond to microscopic regions that are many orders of magnitude smaller than those of astrophysical sources (for example, $\sim10^{-17}$ of the size of relativistic jets).
In  PIC simulations, fast reconnection arises from tearing mode instabilities, producing plasmoids confined to 2D geometries and limited particle acceleration. Accelerated particles can reach energies of up to a few $10^2$ times their rest mass energy ($mc^2$) only and these values are then extrapolated to large scales. The dominant electric field responsible for the acceleration of the particles is resistive, associated with the current density.

 In contrast, 3D MHD simulations incorporating test particles,
are tailored to investigate the macroscopic astrophysical scales of the process. In this regime, fast reconnection is primarily driven by pervasive 3D turbulence, resulting from  violation of the conservation of magnetic flux \citep[][]{lazarian_vishiniac_99, kowal_etal_09, eyink2013, Vicentin2024}. This turbulence can arise from instabilities such as magnetorotational instability (MRI) in accretion disks \citep[][]{Kadowaki_2018} or current-driven kink (CDK) and Kelvin-Helmholtz instabilities
in jets \citep[e.g.][]{kowal_etal_2020, kadowaki_etal_2021, Medina-Torrejón_2021, Medina-Torrejón_2023}.
Turbulent-driven reconnection enables particles to reach ultrahigh energies from the first principles, mainly through a Fermi acceleration mechanism \citep[][]{dalpino_lazarian_2005} facilitated by 3D reconnecting flux tubes at all scales of the turbulent flow \citep[][]{kowal_etal_2012,dalpino_kowal_15,delvalle_etal_16,Medina-Torrejón_2021,kadowaki_etal_2021, Medina-Torrejón_2023}. 
Unlike in PIC scales, the dominant electric field driving this particle acceleration is non-resistive, arising from the ($\mathbf{v} \times \mathbf{B}$) term associated with turbulent magnetic fluctuations in the background flow feeding into the reconnection layers. 

In the Fermi regime, particles are accelerated until reaching a threshold energy, where their Larmor radius matches the turbulence injection scale, which also sets the thickness of the largest reconnection layers. During this regime, their energy grows exponentially in time. Beyond this, particles can gain more energy, but  at a slower rate, due to drift acceleration in large-scale non-reconnecting fields. The resulting energy spectrum typically features a high-energy tail with a power-law index of approximately -1 to -2, shaped by both Fermi and drift mechanisms \citep[][]{kowal_etal_2012, lazarian12, dalpino_kowal_15, delvalle_etal_16, Medina-Torrejón_2021, Medina-Torrejón_2023, dalpino_torrejon2024}. 
This is in contrast with recent 3D PIC simulations that suggest drift acceleration may dominate the particle spectrum over Fermi acceleration \citep[][]{sironi2022, zhang2021, Zhang2023}, but this remains debated \citep[e.g.][]{Guo2023}. As drift acceleration strongly depends on the energy, it is much less efficient at high energies compared to the Fermi mechanism. Therefore, it is unlikely to solely account on drift  for the observed ultra-high energies, underscoring the need for caution when extrapolating PIC results to macroscopic systems \citep[][]{
dalpino_torrejon2024}.

Still, PIC and MHD simulations offer complementary insights. PIC simulations effectively address particle acceleration from sub-rest-mass energies to several hundred times this value, particularly in electron-positron plasmas. 
PIC addresses the injection energy problem.
On the other hand, MHD simulations probe up to the highest observed energies at macroscopic turbulence injection scales, capturing the threshold or saturation energies, particularly focossing on proton acceleration.


Reconnection-driven acceleration offers a potential explanation for the high-energy phenomena observed in AGNs. This talk explores this process within the context of these sources, with focus on the MHD approach.

\section{Reconnection acceleration in blazar jets and the origin of the gamma-rays and neutrinos}

Blazars, which are highly beamed AGN jets pointing to the line of sight, are the most frequent sources of gamma-rays. Theoretically, the particle acceleration mechanisms responsible for the VHE blazar emission and the precise location within the jet where this occurs remain undetermined. 
Recent observations have strongly indicated that these jets can accelerate not only electron-positron pairs but also protons, as exemplified by the TXS 0506+056 source, which exhibited simultaneous emissions of high-energy gamma-rays and neutrinos \citep{neutrinosmm}.

Motivated by findings like this, we have conducted 3D MHD and 3D MHD-PIC simulations to study the transition of relativistic jets from magnetically to kinetically dominated states, with magnetization parameter (given by the ratio between the magnetic and the rest mass energy density) near 1 \citep{Medina-Torrejón_2021,kadowaki_etal_2021,Medina-Torrejón_2023}.  We found that jets destabilized by the current-driven kink instability become turbulent, forming multiple fast reconnection layers across the turbulent flow, consistent with the theory of \cite{lazarian_vishiniac_99}. Test protons injected into this environment experience acceleration mainly due to the Fermi process, with nearly exponential growth in energy in time \citep[][]{dalpino_lazarian_2005}, followed by slower drifting after they escape the reconnection regions. With background magnetic fields around $\sim 10$ G, accelerated particles can attain energies of up to $\sim10^{9} mc^2$ \citep[][]{Medina-Torrejón_2021}. Although these test particle simulations do not account for radiative losses, they provide compelling evidence for these jets' ability to accelerate particles up to VHE energies and produce associated VHE gamma-rays and neutrinos.

Multiple radiative approaches were previously considered in attempts to explain the multi-messenger (MM) emission from TXS 0506+056. Recently, in order to investigate the maximum potential for neutrino production at energy levels consistent with the IC-170922A event, we employed a hybrid lepto-hadronic model without the influence of external soft-photons, assuming  that turbulent-driven magnetic reconnection accelerates protons and electrons by the Fermi process \citep[][]{dalpino2024}.  

\begin{figure}
   \centering
   \includegraphics[width=\hsize]{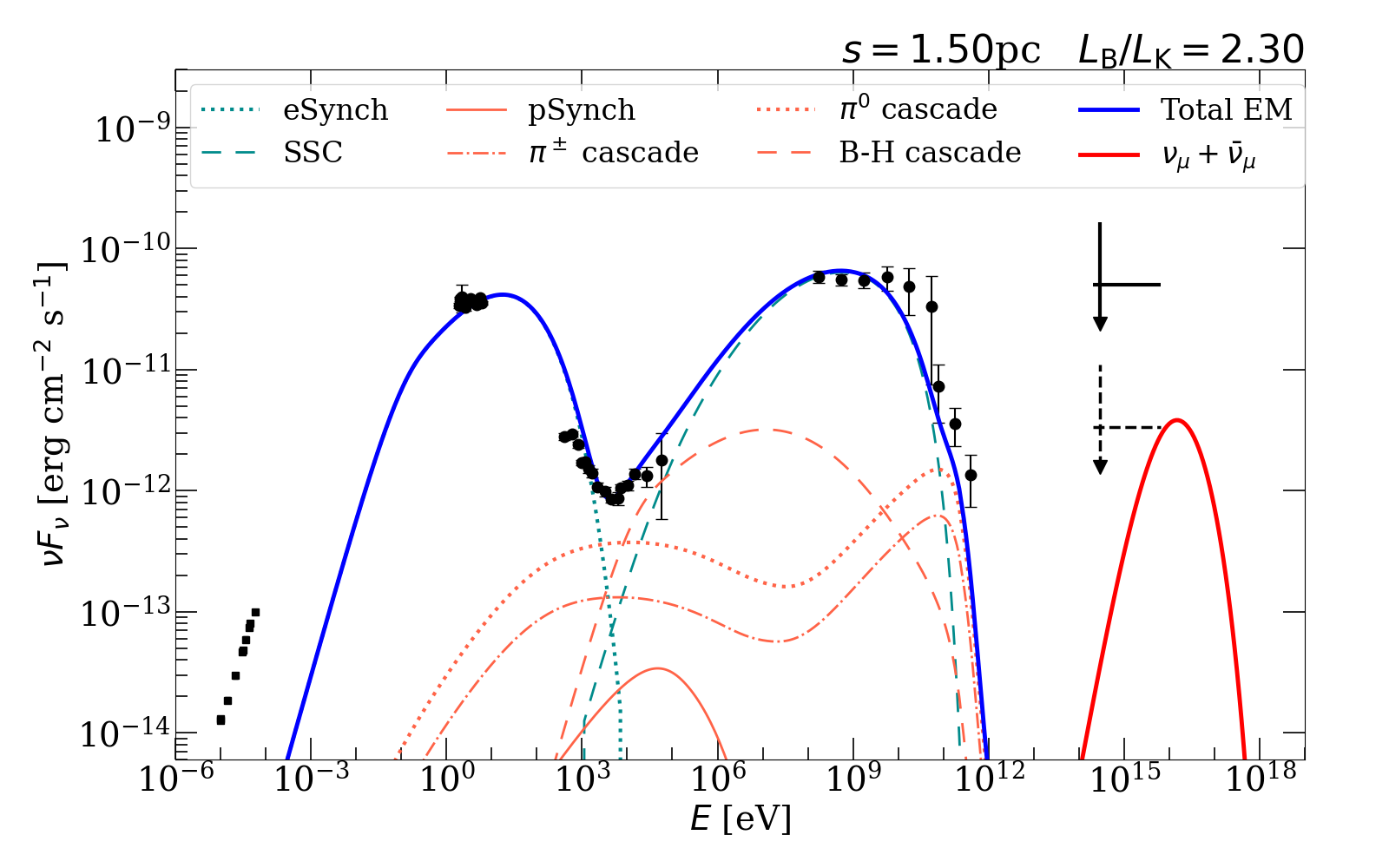} 
      \caption{
MM spectrum of TXS 0506+056 during the 2017 neutrino flare  computed using a lepto-hadronic emission model powered by turbulent-driven magnetic reconnection in the jet. The observed SED data is overlaid, showing consistency with the model. The red curve represents the all-flavor neutrino emission, while the blue curve depicts total emission composed by leptonic (synchrotron and synchrotron-self-Compton, cyan curves) and hadronic (neutral and charged pion, and Bethe-Heitler, crimson curves) contributions. The emission profile is specific to a jet location \( s \), with magnetization $\sigma=L_\mathrm{B}/L_\mathrm{K}$\( \sigma \) derived from the jet model parameters. (Modified from 
 \citep{dalpino2024}).
}
\label{fig:SEDs_1.50pc}
   \end{figure}

We modeled the blazar jet's evolution from a magnetically dominated to a kinetically dominated flow during its propagation. While the simulations described above depict a complex environment with helical and turbulent magnetic fields, we adopted a simplified, but similar framework \citep{2019MNRAS.484.1378G} to provide an analytical description of fast reconnection, magnetic energy dissipation, and particle energization.
We then derived the MM emission at the jet locations where turbulent magnetic reconnection is the operating mechanism of particle acceleration.
With this approach, as the emission region  moves downstream with the jet flow, it produces a sequence of SEDs that are able to reproduce the 2017 MM flare from  TXS-0506+056. We refer to \cite{dalpino2024} for more details, but
as an example, Figure \ref{fig:SEDs_1.50pc} shows one of these SEDs, when the blob is located at a distance $s=1.5$ pc away from the black hole.
We have obtained a time delay between the neutrino and VHE events  $\simeq 6.4$ days, which is  consistent with that observed  in the 2017 MM flare.
We have applied a similar turbulent-driven magnetic reconnection
acceleration/dissipation model to the blazar jet of Mrk 501. During July 2014, multi-wavelength
monitoring by MAGIC of this flaring blazar revealed a TeV gamma-ray spike feature coinciding with a significant enhancement in its X-ray flux. This spectral feature strongly suggests the presence of an extra emission
component beyond the usual one-zone SSC scenario \citep{MRK5012020}. Several theoretical explanations for this novel behavior have been previously proposed, including stochastic particle acceleration, magnetospheric vacuum gap, and pion decay, but all of them failed. We have demonstrated
that the TeV narrow feature simultaneous with an increase in the X-ray flux can be produced by two
leptonic emission regions in a jet undergoing magnetic reconnection energy dissipation. In this scenario, the stable spectral components are produced in the region of maximum magnetic dissipation. A second region, located upstream in the jet in a slower, more magnetized, and
much smaller area compared to the stable one, produces a flare that increases the X-ray flux and generates the mistereous TeV spike (Rodriguez-Ramirez et al. 2025, in prep.).

\section{Reconnection acceleration in accretion flows of AGNs and the origin og gamma-rays and neutrinos}

 Several low-luminosity AGNs (LLAGNs), 
 have been detected at TeV energies \citep[e.g.,][and references therein]{dalpino_etal_2020}.  The poor angular resolution and sensitivity of these detectors make it challenging to determine whether the emission originates from the jet or the core. These detections are surprising due to the underluminous nature of LLAGNs and the moderate Doppler boosting expected from their jets, which are oriented at significant angles to the line of sight. Additionally, short timescale variability in their $\gamma$-ray emission suggests a compact emission region, potentially within the core, prompting alternative particle acceleration models \citep[e.g.,][]{dalpino_etal_2010b}. 

An interesting  study by \citet{kadowaki_etal_15} plotted $\gamma$-ray luminosity versus black hole (BH) mass for over 230 sources, including LLAGNs, our own galactic center SgrA*, black hole binaries (BHBs), blazars, and GRBs, spanning 10 orders of magnitude in mass and power. This revealed two distinct trends: one for blazars and GRBs, associated with Doppler-boosted jet emission, and another for LLAGNs, SgrA* and BHBs, suggesting alternative origins for their $\gamma$-ray emission.

It has been  proposed that fast magnetic reconnection in the black hole  coronal accretion flow  could accelerate particles and produce VHE emission \citep{dalpino_lazarian_2005, dalpino_etal_2010a, dalpino_etal_2010b}. This model, supported by MHD and GR-MHD simulations \citep[e.g.][]{Kadowaki_2018, dalpino_etal_2020}, shows that reconnection between the BH magnetosphere and accretion disk coronal field lines releases magnetic energy,  accelerating the plasma. Analytical estimates of reconnection power align well with observed $\gamma$-ray luminosities of LLAGNs and BHBs \citep[][]{kadowaki_etal_15, singh_etal_15}. 

Detailed modeling of SgrA* accretion flow corona based on 3D GR-MHD simulations combined with radiative transfer and cosmic ray Monte Carlo cascading has demonstrated that reconnection-accelerated particles can reproduce observed TeV upper limits  \citep[][]{rodriguezramires_etal_18}. This process can also produce neutrinos in LLAGNs, potentially contributing to the IceCube's observed diffuse extragalactic neutrino background \citep[][]{Khiali2016}.
Moreover, the recently reported detection of 1–10 TeV neutrinos from the LLAGN seyfert galaxy NGC 1068 \citep{NGC1068}, has brought new challenges to theorists. The absence of TeV-scale gamma-ray emission suggests these neutrinos originate in a dense corona surrounding the supermassive BH which is able to absorve the gamma-rays via pair production. We have  built a lepto-hadronic model based on particle acceleration by turbulent-driven reconnection as described above,  incorporating pair production, pion production, synchrotron radiation, and inverse Compton scattering losses. Preliminary findings align with the observed emission pattern (Passos Reis et al. 2025, in prep.).

This work was partially financed by the São Paulo Research Foundation (FAPESP), Brasil (Grant 2021/02120-0).

\bibliography{bibliography}{}
\bibliographystyle{aasjournal}

\end{document}